\documentclass[12pt]{article}
\usepackage{times}

\usepackage{amsmath}
\usepackage{amssymb}

\usepackage{graphicx}

\usepackage[super,comma,sort&compress]{natbib}
\usepackage{color} 

\usepackage{lineno}
\usepackage{url}
\usepackage{pdfpages}
\usepackage{setspace} 
\usepackage{hyperref}

\hypersetup{
    bookmarks=true,         
    unicode=false,          
    pdftoolbar=true,        
    pdfmenubar=true,        
    pdffitwindow=false,     
    pdfstartview={FitH},    
    pdftitle={My title},    
    pdfauthor={Author},     
    pdfsubject={Subject},   
    pdfcreator={Creator},   
    pdfproducer={Producer}, 
    pdfkeywords={keywords}, 
    pdfnewwindow=true,      
    colorlinks=true,       
    linkcolor=blue,          
    citecolor=blue,        
    filecolor=magenta,      
    urlcolor=cyan           
}

\usepackage[labelfont=bf,labelsep=period,justification=raggedright]{caption}

\makeatletter
\renewcommand{\@biblabel}[1]{\quad#1.}
\makeatother

\date{}

\title{\bf{
To regulate or not: a social dynamics analysis of the race for AI supremacy
} }

\author{The Anh Han$^{1,\star}$, Lu\'is Moniz Pereira $^{2}$,  Francisco C. Santos$^{3,4}$, Tom Lenaerts$^{4,5,\star}$  } 

\begin{document}
\maketitle
{\footnotesize
\noindent
$^{1}$ School of Computing and Digital Technologies,  Teesside University, Middlesbrough, UK TS1 3BA\\
$^{2}$  NOVA Laboratory for Computer Science and Informatics (NOVA LINCS), 
  Universidade Nova de Lisboa, 
        2829-516 Caparica, Portugal \\
$^3$INESC-ID and Instituto Superior Tecnico, Universidade de Lisboa \\
$^4$ Machine Learning Group, Universit{\'e} Libre de Bruxelles, Boulevard du Triomphe CP212, Brussels, Belgium  \\ 
$^5$ Artificial Intelligence Lab, Vrije Universiteit Brussel, Boulevard de la Plaine 2, 1050 Ixelles, Belgium
\\
$^\star$ Corresponding authors: The Anh Han (T.Han@tees.ac.uk) and Tom Lenaerts (Tom.Lenaerts@ulb.ac.be)
}


\newpage
\section*{Abstract} 
Rapid technological advancements in Artificial Intelligence (AI) as well as the growing deployment of intelligent technologies in new application domains are currently driving the competition between businesses, nations and regions. This race for technological supremacy creates a complex ecology of choices that may lead to negative consequences, in particular, when ethical and safety procedures are underestimated or even ignored. As a consequence, different actors are urging to consider both the normative and social impact of these technological advancements.  As there is no easy access to data describing this AI race, theoretical models are necessary  to understand its dynamics, allowing for the identification of when, how and which procedures need to be put in place to favour outcomes beneficial for all. We show that, next to the risks of setbacks and being reprimanded for unsafe behaviour, the time-scale  in which AI supremacy can be achieved plays a crucial role. When this supremacy can be achieved in a short term, those who completely ignore the  safety precautions are bound to win the race  but at a cost to  society, apparently requiring regulatory actions. Our analysis  reveals that blindly imposing regulations may  not have the anticipated effect as only for specific conditions a dilemma arises between what is individually preferred and globally beneficial. Similar observations can be made for the long-term development case. Yet different from the short term situation, certain conditions require the promotion of risk-taking as opposed to compliance to safety regulations in order to improve social welfare.   These results remain robust both when two or several actors are involved in the race and when collective rather than individual  setbacks are produced by risk-taking behaviour.  When defining codes of conduct and regulatory policies for AI, a clear understanding about the time-scale of the race is thus required, as this may induce important non-trivial effects. 
  \\

\noindent \textbf{Keywords:} AI race modelling, safety, game theory, regulation, time-scale. 

\newpage


\section{Introduction}

Interest in AI has exploded in academia and businesses in the last few years.  This excitement is, on the one hand, due to a series of superhuman performances \cite{silver2017mastering,brown2018superhuman,silver2018general,brown2019superhuman,vinuesa2019role} which have been exhibited.  Although successful in highly specialised tasks, these AI success stories appear  in the imagination of the general public as Hollywood-like Artificial General Intelligence (AGI), able to perform a broad set of intellectual tasks while continuously improving itself, generating thus unrealistic expectations and unnecessary fears \cite{cave2019hopes}. 
On the other hand, this excitement is further promoted by political and business leaders alike, for both  anticipate important gains from turning  previously idle data into active assets within business plans \cite{PwC2017}.  All these (un)announced business, societal and  political ambitions indicate that an AI race or bidding war has been triggered \cite{goldai2,cave2018ai,RT2019},  where stake-holders in both private and public sectors are competing to be the first to cross the finish line and hence the  leader in the development and deployment of  powerful, transformative AI \cite{armstrong2016racing,baum2017promotion,bostrom2017strategic,cave2018ai}.   

Irrespectively of the anticipated benefits, many actors have urged for due diligence as i) these AI systems can also be employed for more nefarious activities, e.g. espionage and cyberterrorism \cite{taddeo2018regulate} and ii) whilst attempting to be the first/best,  some ethical consequences as well as safety procedures may be underestimated or even ignored \cite{armstrong2016racing,cave2018ai} (notwithstanding the issue that certain claims about achieving AGI may be overly optimistic or just oversold). These concerns are highlighted by the many letters of scientists against the use of AI in military applications \cite{FLI_letter2015,FLI_signatories_todate}, the blogs of AI experts requesting careful communications \cite{BrookBlog2017} and the proclamations on ethical use of AI in the world \cite{MontrealDec2018,steels2018barcelona,russell2015ethics,jobin2019global}.  

While potential AI disaster scenarios are many \cite{sotala2014responses,armstrong2016racing,pamlin2015global,schubert2019psychology},  the uncertainties in accurately predicting these risks and  outcomes  are high \cite{armstrong2014errors}. As put forward by the Collingridge Dilemma, the impact of a new technology is difficult to predict unless large steps have been taken in its development and it becomes generally adopted \cite{Collingridge1980}.   Sufficient data is therefore not yet available, requiring a modelling approach to grasp what can be expected in a race for AI supremacy (AIS). Models provide dynamic descriptions of the key features  of this race (or parts thereof) allowing one   to understand what outcomes are possible under certain conditions and what may be the effect of policies  that aim to regulate the race. 
This manuscript  focusses on defining a baseline model that discusses when to expect unsafe or safe AI development behaviour and when this is disruptive, i.e. when it harms social welfare.  Subsequently, it can be employed to evaluate the impact of regulatory mechanisms on the behavioural preferences. We resort to the framework of evolutionary game theory \cite{key:Hofbauer1998,key:Sigmund_selfishnes} to address this issue.

Concretely, the model assumes that in order to achieve AIS in a domain $X$, a number of development steps or rounds ($W$) are required. Large-scale surveys and analysis of  AI experts on their beliefs and predictions about progress in AI suggest that the perceived time-scale for AIS is highly diverse across domains and regions \cite{armstrong2014errors,grace2018will}. 
The model therefore aims to capture these different  time-scales of AIS occurrence: When $W$ is small, AIS can be expected to happen in the near future (early AIS regime) while when $W$ is large, AIS will only be achieved far away in time  (late AIS regime). 

Because this is a race, each participant acts by herself during each step in order to reach the target and differs in the speed ($s$) with which she can complete each of the subtasks. The race thus consists of multiple rounds and the fastest participant will reap the benefit ($b$) at each round  when she finishes before the others, winning the ultimate prize ($B\gg b$) once  she carries out the final step achieving AIS in the domain $X$.  When multiple participants reach the end of an intermediate round or the final target at the same time they share the benefits, i.e. $b$ and $B$, respectively. 

In this race, higher $s$ may only be achievable by cutting corners, implying  that some ethical or safety procedures are  ignored. It takes time and effort to comply to precautionary requirements or acquire ethical approvals. Following a safe development process is thus not only more costly, it also results in a slower development speed.  One can therefore  consider that  i) participants in the AI race that act safely (SAFE) pay a cost $c>0$, which is not paid by participants that ignore safety procedures (UNSAFE)  and ii)  the speed of development of UNSAFE participants is faster ($s>1$), compared to the speed of SAFE participants being normalised to $s=1$. So essentially a SAFE player needs $W$ rounds to complete the task, whereas an UNSAFE player will only need $W/s$.  

Yet, UNSAFE strategists may suffer a personal setback or disaster during the race, losing their acquired payoffs.  The risk is personal for UNSAFE players in the current model. Although the threat is greater for the creator \cite{armstrong2016racing,pamlin2015global}, there may also be repercussions for the other participants or society as a whole, a matter discussed in detail in the Supporting Information (SI). As will be shown, this extension  of spreading repercussions does not influence the results discussed in the next sections. 
The probability that the personal setback occurs is denoted by $p_r$ and assumed to increase linearly with the frequency the participant violates the safety precautions. For example, if a participant always plays SAFE then disaster will not occur, given that \[ \left(\frac{|\mathit{UNSAFE}|}{|\mathit{SAFE}|+|\mathit{UNSAFE}|} \right) p_r =0,\] with $|\mathit{UNSAFE}|$ and $|\mathit{SAFE}|$ indicating the number of SAFE and UNSAFE actions respectively.  A participant that only follows safety half of the time  will incur only half of the time the risk  of disaster over all rounds.  

Finally, the model incorporates the possibility that an UNSAFE player is found out at each step of the race, which is an additional risk for UNSAFE players that corresponds to a simple form of regulation. We therefore assume that with some probability $p_{\textit{fo}}$ those playing UNSAFE might be detected and their unsafe behavior disclosed, leading to $0$ payoff in that round. 

Given these different characteristics of the AIS Race (AISR) model, we can now explore which strategies, involving SAFE an UNSAFE actions, are dominant under which conditions, i.e. the parameters defined by this model. Since we resort to evolutionary game theory to answer this question, we consider a population of size  $Z$ in which players engage in a pairwise (or $N$-player) race.  Each player can choose to consistently follow safety precautions (denoted by \textbf{AS}, the SAFE players)  or completely ignore them (denoted by \textbf{AU}, the UNSAFE players). Additionally,
we assume that, upon realising that UNSAFE players ignore safety precautions to gain a greater development speed,  leading to the wining of the prize $B$ (and a larger  share of the intermediate benefit in each round, $b$, especially in the regime of weak monitoring or low $p_{\textit{fo}}$), SAFE players might adopt unsafeness as well to avoid further disadvantage. It is indeed  observed that competing countries or companies might engage  in such a  safety corner-cutting behaviour in deploying unsafe AI to avoid falling behind \cite{RT2019}.    We therefore consider, in line with previous literature on repeated games  \cite{key:axelrod84,key:Sigmund_selfishnes,key:hanetalAdaptiveBeh,van2012emergence}, a conditional strategy (denoted by \textbf{CS}), which plays SAFE in the first round and then adopts the move its co-player used in the previous round.  This so-called direct reciprocity strategy has been shown to promote cooperation in the context of repeated social dilemmas, outperforming consistently defective individuals \cite{key:axelrod84,key:Sigmund_selfishnes}. Alternative strategies can be imagined but for the sake of simplicity we focus (for now) on these three.

In the following, we will examine, across different time-scales of the AISR, under which conditions (for instance, regarding the disaster probability), safety behaviour should be promoted or externally enforced. Similarly, we address when one should  omit the safety precautions for a larger social welfare to arise faster, when the benefits gained in doing so exceed the risk of a setback or personal disaster.  Moreover, given the first-mover advantage of UNSAFE players in the race to AI supremacy (i.e., acquire $B$), we will examine whether (and under what time-scale of the AISR model) conditional behaviours can still  act as a promoting mechanism to achieve safety when required, or otherwise other  mechanisms are needed. 
For the sake of clarity, we investigate here the pairwise race model and perform the analysis for the  $N$-player ($N \geq 2$) AISR in SI. Additionally, the situation where the effects of a setback or disaster are no longer just personal are analysed in depth in SI.

\section{ Results}

\begin{figure*}
\centering
\includegraphics[width=\linewidth]{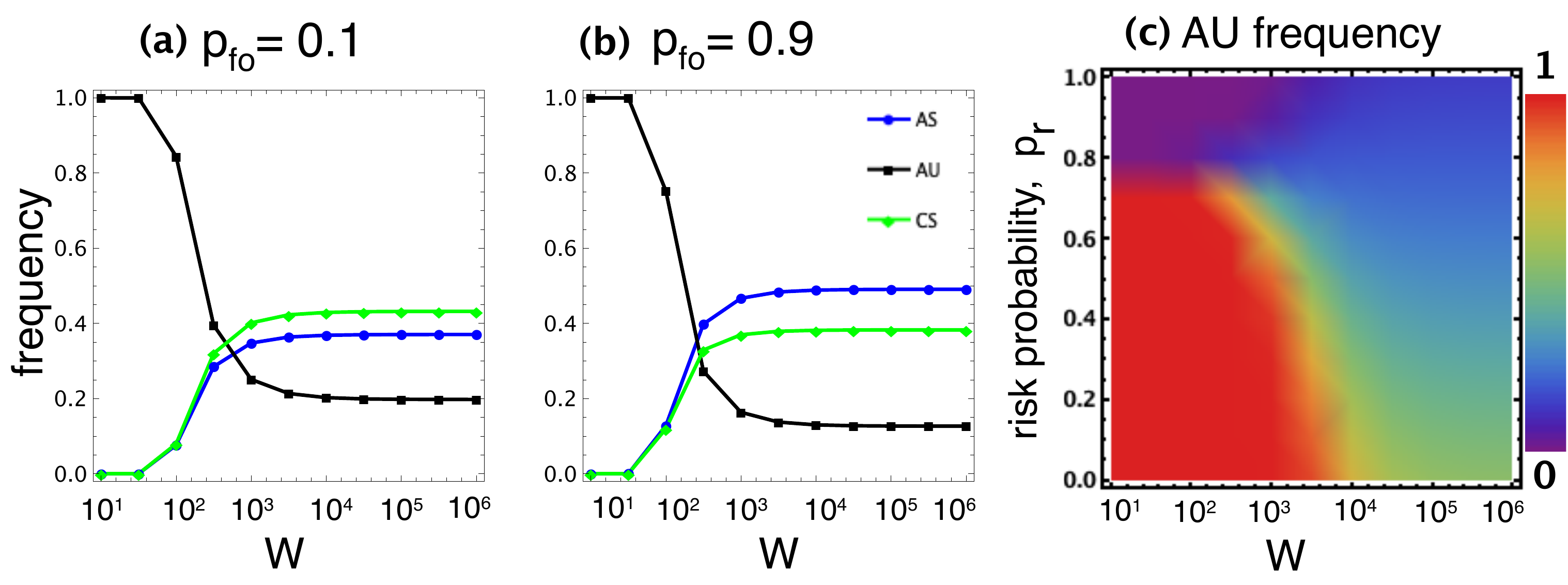}
\caption{ \textbf{Different regimes of AIS: when $W$ is small (early AIS) vs when $W$ is larger (late AIS).} 
Panels (a) and (b) show the frequency of each strategy, i.e.  AS, AU and CS, in a population ($p_r = 0.6$).  In the early AIS regime, AU dominates the population, while AS and CS outperform AU in the late  AIS regime. The former observation is valid for  $p_r$ values lower than $0.8$, see panel (c) ($p_\textit{fo} = 0.1$). For a high risk probability of disaster occurring due to ignoring safety precautions (high $p_r$), AU disappears in both regimes.  The black line in (c) indicates the threshold  of $p_r$ above which SAFE is the preferred collective action and below which UNSAFE is the preferred one. Parameters: $c = 1$, $b = 4$, $s = 1.5$, $B = 10^4$, $\beta = 0.1$,  $Z = 100$. 
}
\label{fig:different_regimes_AIG}
\end{figure*}

 We calculate the long-term frequency  of each possible behavioural composition of the population, the so-called stationary distribution (cf. Methods), as this will reveal the action preferences (i.e. behaving safely or not) of a finite set of virtual players within the context of the AISR game defined above.  
 This stochastic social dynamics of the population occurs in the presence of errors, both in terms of errors of imitation and of behavioural changes, the latter representing  an open  exploration of the possible strategies by the virtual participants \cite{key:Hofbauer1998,key:Sigmund_selfishnes}.
As can be observed in Figure \ref{fig:different_regimes_AIG}, the preference for the strategies AS, AU and CS changes for different lengths of the race.  We distinguish two regimes in the AISR that depend on the relationship between the number of rounds $W$ needed to achieve the ultimate benefit $B$ and the revenue that can be achieved at every round, i.e. $b$:
\begin{itemize}
\item[i)] \textbf{Early AIS}: This regime is characterised by the observation that the ultimate prize of winning the race in $W$ rounds strongly outweighs the benefits that can be achieved in a single round, i.e. $B/W \gg b$. Being fast is thus a key driver here. 
\item[ii)] \textbf{Late AIS}: In this regime, AIS will not be achieved in a foreseeable future, making the gains at each round $b$, even when having to  pay the safety cost $c$, more attractive than  the ultimate prize of winning the race $B$, i.e. $B/W \ll b$.
\end{itemize}
We observe that in the first AIS regime, AU dominates the population whenever   the probability that an AI disaster occurs due to unsafe development ($p_r$) is not too high (see Figure \ref{fig:different_regimes_AIG}c; also panels a and b, where  $p_r = 0.6$).  In the second AIS regime, AS and CS  take over (Figure \ref{fig:different_regimes_AIG}a-b). 
When an AI disaster is more likely to occur due to unsafe developments (i.e. large $p_r$, see Figure \ref{fig:different_regimes_AIG}c), AU disappears in both regimes.

Given the difference in  behavioural preferences toward safety developments in the early and late regimes, different kinds of regulation may be required. 
Since AI developments should at least provide a beneficial outcome for the individual developers and interested users in society, we first investigate under which conditions they can achieve their ambitions by acting safely, thus avoiding the risk of personal setbacks or shared disaster (see SI). When the benefits of all making safe developments ($\Pi_{AS,AS}$) outweigh the benefits of all doing things unsafely ($\Pi_{AU, AU}$), i.e. when $\Pi_{AS,AS} > \Pi_{AU, AU}$, this goal can be achieved (see Methods). The black line in Figure \ref{fig:different_regimes_AIG}c depicts this  threshold in function of $p_r$,  revealing that there is a large part in the early regime (red area above the black line) where regulation should be put in place to restrain unsafe development behaviour. On the other hand, in the late regime (beyond $10^4$ development steps),   risk-taking should be promoted as this will  improve social welfare (area below the black line).  

Figure \ref{fig:different_regimes_AIG} thus underlines the importance of knowing in which regime the race is operating, since  this would affect the type of regulation that one should introduce.   In order to assess these observations in detail, we carry out a more in-depth analysis in the following sections.

\subsection*{Early AIS: only under specific conditions will regulation improve welfare}

We first focus again on the analytical conditions under which $\Pi_{AS,AS} > \Pi_{AU, AU}$ and then determine when the safe and reciprocal strategies are more likely to be imitated as this shows what behaviour to expect when participants can alter their actions in function of the benefits they can gain. 

In the current AIS regime, the first condition occurs when (see SI for the proof)
\begin{equation} 
\label{eq:safety_prefer}
 p_r > 1 - \frac{1}{s}. \ \quad 
\end{equation}
That is,  when the risk  of a personal setback ($p_r$)  is larger than  the gain one can get from a greater development speed, then safe development is the preferred collective action in the population, and vice versa.  

\begin{figure*}
\centering
\includegraphics[width=\linewidth]{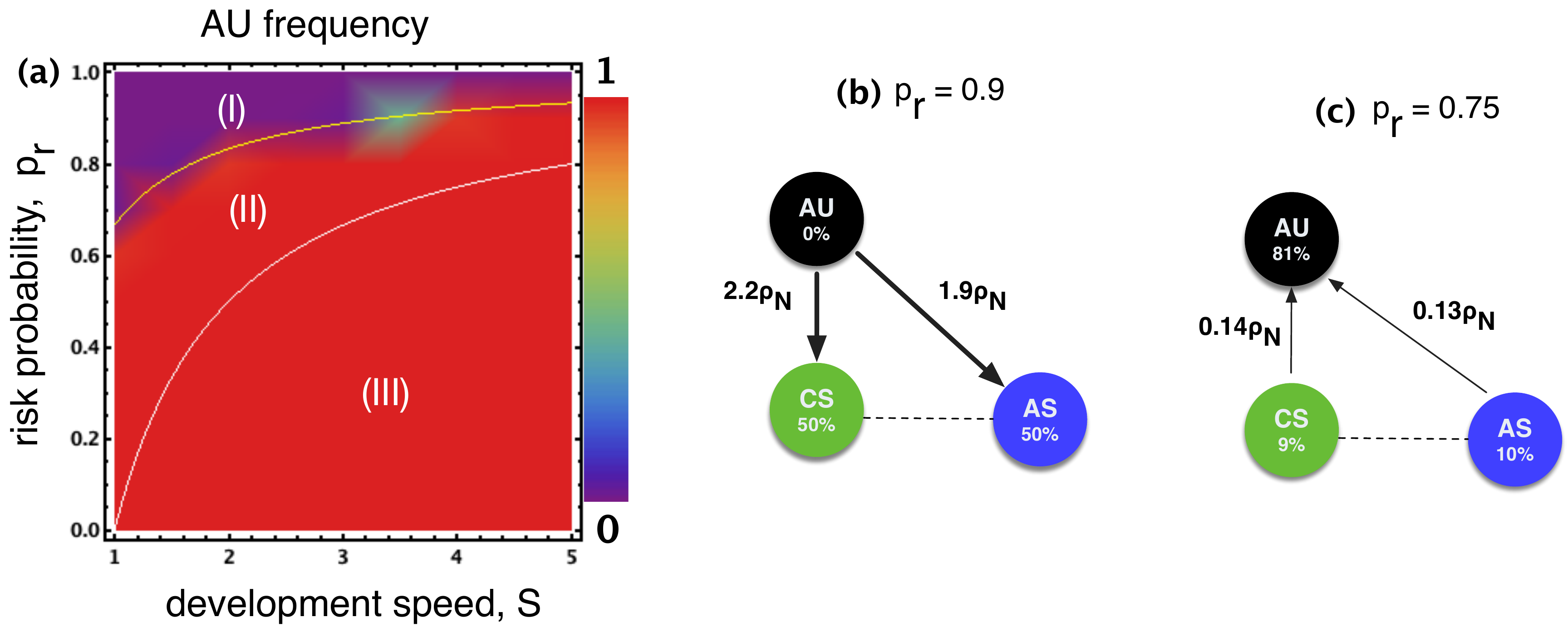}
\caption{\textbf{Early AIS regime}.   \textbf{(a)} Frequency of AU as a function of the speed gained, $s$, and the probability of AI disaster occurring,  $p_r$, when ignoring safety. In general, we observe that when the risk probability is small, AU is dominant. The larger $s$ is,  AU dominates for a larger range of $p_r$.  Region (\textbf{II}): The two solid lines inside the plots indicate the boundaries $p_r \in [1-1/s, 1-1/(3s)]$  where safety development is the preferred collective outcome but unsafe development is selected by social dynamics. Regions (\textbf{I}) (resp., (\textbf{III})) indicate where safe (resp., unsafe) development is both the preferred collective outcome and the one selected by social dynamics.  Panels \textbf{(b)} ($p_r = 0.9$)  and \textbf{(c)} ($p_r = 0.6$): transition probabilities and stationary distribution in a population of AS, AU, and CS, with $s = 1.5$.  AU dominates in panel (c), corresponding to region (\textbf{II}), while AS and CS dominate in panel (b), corresponding to region (\textbf{I}). We only show the stronger directions.  Parameters: $c = 1$, $b = 4$, $W = 100$, $p_{\textit{fo}} = 0.5$, $B = 10^4$, $\beta = 0.1$,  $Z = 100$. 
}
\label{fig:panel_no_punishment}
\end{figure*}
Analysis of the second question, i.e. when safe (AS) and conditionally safe (CS) strategies are more likely to be imitated, reveals that both  are preferred over AU by the social learning dynamics we use here  (see risk-dominance analysis in SI) when   
\begin{equation}
\label{eq:risk_dom_AUvsCSAS}
p_r > 1-\frac{1}{3s}.
\end{equation}
The two boundary conditions  in Equations  \ref{eq:safety_prefer} and \ref{eq:risk_dom_AUvsCSAS} divide the space defined by the speed of development ($s$) and the risk of disaster ($p_r$) into  three regions, as shown in Figure \ref{fig:panel_no_punishment}a: 
\begin{itemize}
\item[(\textbf{I})]  when $ p_r > 1-\frac{1}{3s}$:  This is the \emph{AIS compliance zone}, where safe AI development is both the preferred collective outcome and fully safe or conditionally safe behaviour is the social norm (see Figure \ref{fig:panel_no_punishment}b for an example: for $s = 1.5$ the condition becomes $p_r > 0.78$); 
\item[(\textbf{II})] when $1-\frac{1}{3s} > p_r > 1-\frac{1}{s}$: This intermediate zone captures a dilemma since, collectively, safe AI developments are preferred, yet the social dynamics pushes the population to the state where everyone develops AI in an unsafe manner. We will refer to this zone as the \emph{AIS dilemma zone} (see Figure \ref{fig:panel_no_punishment}c for an example:  for $s = 1.5$ the condition becomes $0.78 > p_r > 0.33$);  
\item[(\textbf{III})] when $p_r <  1-\frac{1}{s}$: This is the \emph{AIS innovation zone}, where  unsafe development is both the preferred collective outcome and the one selected by the social dynamics.  
\end{itemize}


The results visualised in Figure \ref{fig:panel_no_punishment} remain present for different parameter settings as is shown in Figure S4 in the SI. 


As can be observed, in regions (\textbf{I}) and (\textbf{III}), the preferred collective outcomes are also selected by the social dynamics.  Whereas in the AIS compliance zone, the high risk of disaster motivates participants to adopt a safe strategy even when the final benefit $B$ outweighs marginal benefits per round. In the latter, the AIS innovation zone, the benefit of quickly reaching AIS is everything and speed ensures that one arrives first, with limited risk for a setback or even shared disaster (see SI). In terms of social welfare, i.e. the average benefits spread over the population, the AIS innovation zone  produces the largest benefits, especially for low risk and  high speed combinations (see SI,  Figure S13).  In the AIS compliance zone, the social welfare is stable no matter the speed, yet lower than in (\textbf{III}). Yet switching  to unsafe actions here would only lead to a worse outcome, so compliance to safety and ethical regulations are thus required. 

Region (\textbf{II}), the AIS dilemma zone, is somewhat peculiar as collective safe behaviour is preferred, yet social dynamics selects for unsafe behaviour.  As a consequence, social welfare is lower than what can be seen in the two other zones.  Regulation of unsafe behaviour is thus required here as it will nudge the social dynamics towards safe behaviour and, consequently, greater overall social welfare. Such regulation activities will have no effect in the AIS compliance zone and are potentially detrimental (in terms of the missed social welfare) effects in the AIS innovation zone. It is therefore essential to know, when the time-scale to reach AIS is short, what risks can be expected and what speed is acceptable to avoid the AIS dilemma zone and ensure a positive effect for society.    

Looking back at the observation in Figure \ref{fig:different_regimes_AIG} that in the early AIS regulation is necessary, the current analysis reveals that this is only a necessity when risk and development speed put the race in the AIS dilemma zone since the effects would be counterproductive in the two other zones. Yet stimuli to promote risk-taking in the AIS innovation zone and following safety protocols in the AIS compliance zone are potentially useful when participants in the race are unsure about the importance of following those actions, i.e. when participants are still exploring and not imitating enough the most beneficial behaviours --- expressed by imitation strength $\beta$ in our model (cf. Figure S4 in SI) ---  in those zones.



Note that the boundaries established by Equations  \ref{eq:safety_prefer} and \ref{eq:risk_dom_AUvsCSAS}  are applicable for both CS and AS when playing against AU. Thus, similar results are obtained  if we consider  a population of just two strategies AS and AU (cf. Figure S5 in SI). Adding CS does not change the overall outcome and conditions for safe AI development to be selected.  These results also remain unchanged when the risk of setbacks  is not just personal, i.e. being shared among the race participants (whether equally or not), as shown  analytically in SI (also see Figure S10). The results are furthermore robust to changes in the number of participants in the race. When considering the AI race among $N$ development teams (see SI), the main difference is that the upper bound of region (\textbf{II}) increases. That is, the AIS dilemma zone increases and the AIS compliance zone disappears. Regulation is thus required for a larger part of the speed-disaster space (cf. Figures S7 and S8 in SI). The reason is, the larger the group size the greater the chance that there is at least one AU player in the group with other AS and CS players, who would then win the development race.


\subsection*{Late AIS: risk-taking as opposed to safety compliance may need to be promoted}


\begin{figure*}
\centering
\includegraphics[width=\linewidth]{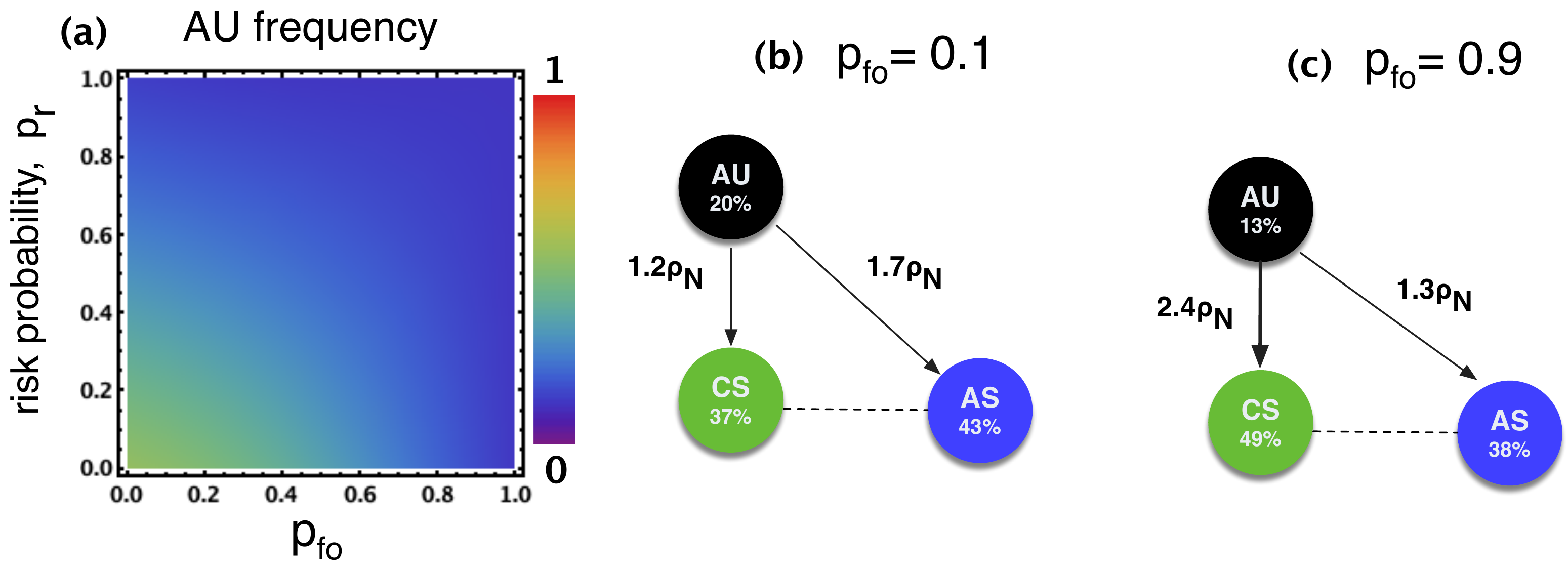}
\caption{\textbf{Late AIS regime}. 
\textbf{(a)}  Frequency of AU as a function of the probability of unsafe development being found out, $p_\mathit{fo}$, and the probability of AI disaster occurring  $p_r$, when the number of development steps to reach AIS is  large ($W = 10^6$). AU has a low frequency whenever  $p_\mathit{fo}$ or $p_r$ are sufficiently high. The lines  indicate the  conditions above which  safety behavior is the preferred collective outcome (black line) and when AS and CS are risk-dominant against AU (blue and green lines, respectively). CS is risk-dominant for a larger range   of $p_r$ than AS for small $p_\mathit{fo}$, which is reversed for large  $p_\mathit{fo}$. The numbers refer again to the three zones, i.e. the AIS compliance, the AIS dilemma and the AIS innovation zones.  \textbf{(b-c)}: transition probabilities and stationary distribution ($p_r = 0.4$). Against AU, AS performs better than CS   when  $p_\mathit{fo}$ is large, which is reversed when     $p_\mathit{fo}$ is  small.   Parameters: $c = 1$, $b = 4$, $s=1.5$,  $B = 10^4$, $\beta = 0.1$, $Z = 100$. 
}
\label{fig:panel_freq_largeW}
\end{figure*}

When AIS is unachievable in the short term,  AS and CS are the dominant social norms,   as was shown in Figure \ref{fig:different_regimes_AIG}. However, when the probability of  disaster is rather small, unsafe behaviour would lead to a relatively greater welfare, yet overall much less than in the early AIS regime (see SI). In Figure \ref{fig:panel_freq_largeW}, one can again distinguish three zones, i.e. the AIS compliance, AIS dilemma and AIS innovation zones, based on conditions for which safety behaviour is the preferred collective outcome  and when AS and CS are risk-dominant against AU (see the black, blue and green lines, respectively, in Figure \ref{fig:panel_freq_largeW}).


In both the late AIS compliance and late AIS innovation zones, regulation is not required as before. Although, as also pointed out in the previous section, stimulating a faster acquisition of the required behaviour in those zones can potentially be useful. In the late AIS dilemma zone, regulation should be put in place to enforce behaviour that improves social welfare. However, in contrast to the early AIS where safety should be promoted, in this late AIS regime, unsafe behaviour (speedy innovation) should be promoted to increase social welfare (see Figure S14 in SI). This zone covers the area in-between intermediate $p_r$ with low $p_\textit{fo}$, and low $p_r$ with intermediate $p_\textit{fo}$. In both areas decreasing the level of monitoring leads to better social welfare. In the latter  where $p_r$ is low, decreasing $p_\textit{fo}$ would move it into the innovation zone. In the former, despite not completely removing the dilemma, decreasing $p_\textit{fo}$ increases the frequency of AU and the overall social welfare.    Interestingly, high levels of detection risk removes the dilemma zone, moving both areas into the compliance zone, as also can be observed in  Figures S1 and S2 in SI for other parameter settings, yet lower social welfare is obtained. Note however that in the compliance zone where $p_r$ is high, social welfare is highest for intermediate levels of monitoring (see Figure S15 in SI). 

As shown in the SI,  the  observations remain valid if, instead of pairwise interactions, one considers a race with $N>2$ teams in the late AIS regime, i.e. all three zones reappear. Moreover, when $N$ increases, while the innovation zone size remains unchanged, the AIS dilemma zone again increases. Also in this case AU becomes the preferred collective outcome for a wider range of $p_r$ and $p_{fo}$ (see Figure S9 in SI). 
Additionally, when the risk of disaster is not just personal  but is rather  shared among the race participants, we observe that the preference boundary between collectively safe and unsafe behaviour remains the same yet the individual preference towards risky development  increases, i.e. the innovation zone becomes larger while the dilemma zone becomes smaller and disappears (see Figure S11 in SI). That is, shared risk in the late AIS regime improves the overall social welfare (by allowing more beneficial innovation to happen), reducing  the need for regulatory actions to handle the late AIS dilemma zone.

\section{Discussion} 

Our results  reveal that knowing the exact timing of reaching AIS in a domain is not crucial, only whether it can be achieved  early or late, as this will influence what regulations are potentially suitable.  We identified three different AIS zones in both the early and late regimes, i.e. the safety compliance, the dilemma  and the innovation zones. They are respectively characterised by high risk, intermediate risk and low risk for personal as well as shared setbacks. In the compliance and innovation zones,  regulatory actions that reverse the behaviour selected by social dynamics should be avoided, as they would be detrimental to the overall social welfare. Stimulating, on the other hand, a faster acquisition of the required behaviour in those zones can potentially be useful. In the dilemma zone, however,   regulatory actions promoting the collectively beneficial outcome are essential since the behaviour selected by social dynamics goes against society's interest, lowering social welfare. In this AIS zone the social dynamics is selecting for (undesired) behaviour, requiring regulation of  risk-taking in the early AIS and safety compliance in the late AIS. 
 

We show furthermore, both in the early and late regimes, that although the three AIS zones are determined by similar ranges of the risk for  setbacks  ($p_r$), they  differ in the secondary factors that control the extent of these zones. While in the early AIS, speedy development ($s$) is everything, the race outcome in the late AIS is mainly determined by the efficiency and level of monitoring of unsafe behaviour ($p_\textit{fo}$). Although speed in the early regime appears to handle some levels of disaster risk,  it may lead participants to enter the dilemma zone where individual interests counter societal welfare, and this area increases in function of the number of participants in the race.  As is shown in Figure S2, speed does not influence the regions in the late AIS regime. The risk of being detected actually limits unsafe behaviour to the area of low risk situations. Yet more participants will increase again the area (see Figure S9) as well as sharing the effects of a disaster (see Figure S11).  It appears thus that holding unsafe players responsible for bad outcomes of the AIS race will ensure, at least in the late regime, that unsafe actions remain limited. Moreover, the presence of conditionally safe players, i.e. the threat that others may also start behaving unsafely,  limits the unsafe actions to lower risk areas.  

The AISR model and associated analysis provides thus an instrument for policy makers to think about the supporting mechanisms (such as suitable rewards and sanctions) \cite{key:Sigmund_selfishnes,sotala2014responses,szolnoki2013correlation,Han:2014tl,hanAIES2019,vinuesa2019role} needed to mediate a given AI race. In the early AIS, controlling  the  development speed of AI teams appears essential. Yet, policy makers should carefully consider whether it will have the expected outcome, i.e. whether the race is actually occurring in the AIS dilemma zone. 
In the late AIS,   monitoring was perceived to be crucial. Decreasing the level of monitoring  can reduce the  dilemma zone and increases  social welfare, increasing speedy innovation. Intermediate levels of monitoring  lead to highest social welfare in the compliance zone.


Moreover, one should consider  the possibility that the risk of  being identified as an unsafe player may not just affect a single development round, but may also have repercussions on subsequent rounds, i.e. the unsafe player may also loose $b$ for instance in all  subsequent rounds. As shown in  SI the results remain the same in early AIS, while in the late AIS, the outcome is equivalent to the results one obtains when full monitoring (i.e. $p_\textit{fo} = 1$) is in effect. Intuitively,    longer consequence associated with being detected is equivalent to having a higher probability of being detected in each round in the current AISR model.



There are of course limitations to the current model, which will require further analysis. On the one hand,  the effect of unsafe behavior on $W$ has not been considered. It may well be that accumulated detected unsafe behaviour, whether by a single player or jointly accumulated by a number of them, may expand the time necessary to reach the AIS, thus effectively increasing $W$. Moreover, the time to reach AIS in a domain $W$ may also be affected by the trust that people have in  AI techniques, even when deliberate unsafe behaviour is not the issue. 
Rhetoric and framing of the AI development race and how close it is to achieve the AGI might strongly influence the dynamics and outcome of the AI race \cite{cave2018ai,baum2017promotion}.  In future work, such phenomena should be examined and introduced on top of the base model presented.

One the other hand, the model also did not consider that to achieve AIS in some domain, the results of multiple races may not to be combined. Here long-term targets like AGI are considered to be achievable in one race.  Clearly AGI will require solutions to multiple subproblems, which by themselves may be achieved in development efforts occurring at different time scales.  Future models of AISR will thus need to consider that multiple AISR games to study what regulatory actions are most beneficial for this kind of goals. 

Notwithstanding all additional features one can imagine that are interesting for framing the AISR, the current work provides a thoroughly  analysed base-line model that can be used to answer  relevant questions on the regulation of innovation and research activities in the current races for different kinds of AI supremacy.  

In conclusion, we have provided here a first plausible AISR model directly useful for policy makers and researchers to evaluate the risks associated with the ongoing AI development and applications race, and have shown and analysed its reasonably acceptable behavioural consequences. Our results indicate the crucial need of clarifying the time-scale of digital innovation supremacy and the  risks in relation to ignoring safety and ethical precautions in speeding up innovation,  in order to determine suitable regulations of  AI safety behaviour beneficial for all.

\newpage 
\section{Methods}
\paragraph{AI race model definition.}

The AI development race is modeled as a repeated two-player game, consisting of $W$ development rounds. In each round, the players can collect benefits from their intermediate AI products, depending on whether they choose to play SAFE or UNSAFE. Assuming a fixed benefit, $b$,  from the AI market, teams  will share this benefit proportionally to their development speed. Moreover, we assume that with some probability $p_{\textit{fo}}$ those playing UNSAFE might be found out about their unsafe development and their products won't be used, leading to 0 benefit.  
Thus, in each round of the race, we can write the payoff matrix as follows (with respect to the row player) 
{\small
\begin{equation}
\pi =  \bordermatrix{~ & \textit{SAFE} & \textit{UNSAFE}\cr
                  \textit{SAFE} & -c + \frac{b}{2} &-c + (1-p_{\textit{fo}}) \frac{b}{s+1}    + p_{\mathit{fo}} b  \cr
                  \textit{UNSAFE} & (1-p_{\textit{fo}}) \frac{s b}{s+1}   & (1-p^2_{\textit{fo}}) \frac{ b}{2}   \cr
                 }.
\end{equation}
}
For instance, when two SAFE players interact, each needs to pay the cost $c$ and they share the benefit $b$. 
When a SAFE player interacts with an UNSAFE one the SAFE player pays a cost $c$ and  obtains the full benefit $b$ 
in case the UNSAFE co-player is  found out (with probability $p_{fo}$), and obtains a small part of the benefit $b/(s+1)$ otherwise 
(i.e. with probability $1 - p_{fo}$).  When playing with a SAFE player,  the UNSAFE does not have to pay any cost and obtains 
a larger share $bs/(s+1)$ when not  found out. Finally, when an UNSAFE player interacts with another UNSAFE, it 
obtains the shared benefit $b/2$ when both are not found out and the full benefit $b$ when it is not found out while the co-player is found out, and 0 otherwise. The payoff is thus:  $(1-p_{fo})\left[  (1-p_{fo}) (b/2) + p_{fo} b \right] = (1-p^2_{fo}) \frac{ b}{2}.$ 
The payoff matrix defining averaged payoffs for the three strategies reads  
{\small
 \begin{equation}
 \label{eq:AVERAGE-PAYOFF-AIRACE}
\Pi = \bordermatrix{~ & \textit{AS} & \textit{AU} & \textit{CS} \cr
                  \textit{AS} & \frac{B}{2W} +\pi_{11} & \pi_{12}   &  \frac{B}{2W}+\pi_{11} \cr
                  \textit{AU} &  (1 - p_r) \left(\frac{sB}{W} + \pi_{21}\right)   &  (1 - p_r) \left(\frac{sB}{2W} +\pi_{22}\right)  &  (1 - p_r)\left[\frac{sB}{W} +  \frac{s }{W}  \left(\pi_{21} + (\frac{W}{s} - 1) \pi_{22}  \right)\right] \cr 
                  \textit{CS} &  \frac{B}{2W} +\pi_{11}    & \frac{s}{W}  \left( \pi_{12} + (\frac{W}{s} - 1) \pi_{22}  \right) &  \frac{B}{2W} +\pi_{11}    \cr
                 }.
\end{equation}}

\paragraph{Evolutionary Dynamics in Finite Populations.}
We adopt here  evolutionary game theory (EGT) methods for finite populations to derive analytical results and numerical observations \cite{key:novaknature2004,key:imhof2005,nowak:2006bo}.  In a repeated games,  players'  average payoff over all the game rounds (see the payoff matrix in Equation \ref{eq:AVERAGE-PAYOFF-AIRACE}) represents their \emph{fitness} or social \emph{success}, and  evolutionary dynamics is shaped  by social learning \cite{key:Hofbauer1998,key:Sigmund_selfishnes}, whereby the  most successful players will tend to be imitated more often by the other players. In the current work, social learning is modeled using  the so-called pairwise comparison rule \cite{traulsen2006},  assuming  that a player $A$ with fitness $f_A$ adopts the strategy of another player $B$ with fitness $f_B$ with probability given by the Fermi function, 
$\left(1 + e^{-\beta(f_B-f_A)}\right)^{-1}$,  
where  $\beta$ conveniently describes the selection intensity ($\beta=0$ represents neutral drift while $\beta \rightarrow \infty$ represents increasingly deterministic selection). 
 For convenience of numerical computations, but without affecting analytical results, we assume here small mutation limit\cite{Fudenberg2005,key:imhof2005,key:novaknature2004}. 
As such,  at most two strategies are present in the population simultaneously, and  the behavioural dynamics can thus be  described by a Markov Chain, where each state represents a homogeneous population and  the transition probabilities between any two states  are given by the fixation probability of a single mutant \cite{Fudenberg2005,key:imhof2005,key:novaknature2004}. The resulting Markov Chain has a stationary distribution, which describes the average time the population spends in an  end state. In two-player game, the average payoffs in a population of  $k$ A players  and $(Z-k)$ B players can be given  as below (recall that  $Z$ is the  population size), respectively, 
\begin{equation} 
\label{eq:PayoffA}
\begin{split} 
P_A(k) =\frac{(k-1)\Pi_{A,A} + (Z-k)\Pi_{A,B}}{Z-1},  \quad P_B(k) =\frac{k\Pi_{B,A} + (Z-k-1)\Pi_{B,B}}{Z-1}.
\end{split}
\end{equation}
The fixation probability that a single mutant  A taking over a whole population with  $(Z-1)$ B players is as follows \cite{traulsen2006,Karlin:book:1975, key:novaknature2004}
\begin{equation} 
\label{eq:fixprob} 
\rho_{B,A} = \left(1 + \sum_{i = 1}^{Z-1} \prod_{j = 1}^i \frac{T^-(j)}{T^+(j)}\right)^{-1},
\end{equation} 
where $T^{\pm}(k) =  \frac{Z-k}{Z} \frac{k}{Z} \left[1 + e^{\mp\beta[P_A(k) - P_B(k)]}\right]^{-1}$ describes  the probability to change the number of A players  by $\pm$ one in a time step.
Specifically, when $\beta = 0$, $\rho_{B,A} = 1/Z$, representing the transition probability at neutral limit. 

Having obtained the fixation probabilities between any two states of a Markov chain, we  can  now describe its stationary distribution \cite{Fudenberg2005,key:imhof2005}. Namely, considering a set of $s$ strategies,  $\{1,...,s\}$, their stationary distribution  is given by the normalised eigenvector associated with the eigenvalue $1$ of the transposed of  a matrix $M = \{T_{ij}\}_{i,j = 1}^s$, where $T_{ij, j \neq i} = \rho_{ji}/(s-1)$ and  $T_{ii} = 1 - \sum^{s}_{j = 1, j \neq i} T_{ij}$. 

\paragraph{Risk-dominant conditions.} We can determine  which selection direction is more probable: an A mutant  fixating in a homogeneous population of individuals playing B or a B mutant fixating in a homogeneous population of individuals playing A. When the first is more likely than the latter, A is said to be \emph{risk-dominant} against B \citep{kandori:1993aa,GokhalePNAS2010}, which holds for any intensity of selection and in the limit of large $N$ when 
\begin{equation} 
\label{eq:compare_fixprob_cond_largeN}
\pi_{A,A} + \pi_{A,B} > \pi_{B,A} + \pi_{B,B}.
\end{equation}

\newpage
\section{Acknowledgements}
T.A.H., L.M.P. and T.L. are supported by Future of Life Institute grant RFP2-154. L.M.P. acknowledges support from FCT/MEC NOVA LINCS PEst UID/CEC/04516/2019. F.C.S. acknowledges support from FCT Portugal (grants PTDC/EEI-SII/5081/2014, PTDC/MAT/STA/3358/2014). T.L. acknowledges support by the FuturICT2.0 (www.futurict2.eu) project funded by the FLAG-ERA JCT 2016.

%

%

\begin{thebibliography}{10}

\bibitem{goldai2}
AI-Roadmap-Institute.
\newblock Report from the ai race avoidance workshop, tokyo.
\newblock 2017.

\bibitem{RT2019}
Peter Apps.
\newblock {Are China, Russia winning the AI arms race?}, January 2019.
\newblock [Reuters; Online posted 15-January-2019].

\bibitem{armstrong2016racing}
Stuart Armstrong, Nick Bostrom, and Carl Shulman.
\newblock Racing to the precipice: a model of artificial intelligence
  development.
\newblock {\em AI \& society}, 31(2):201--206, 2016.

\bibitem{armstrong2014errors}
Stuart Armstrong, Kaj Sotala, and Se{\'a}n~S {\'O}~h{\'E}igeartaigh.
\newblock The errors, insights and lessons of famous ai predictions--and what
  they mean for the future.
\newblock {\em Journal of Experimental \& Theoretical Artificial Intelligence},
  26(3):317--342, 2014.

\bibitem{key:axelrod84}
Robert Axelrod.
\newblock {\em The Evolution of Cooperation}.
\newblock Basic Books, ISBN 0-465-02122-2, 1984.

\bibitem{baum2017promotion}
Seth~D Baum.
\newblock On the promotion of safe and socially beneficial artificial
  intelligence.
\newblock {\em AI \& SOCIETY}, 32(4):543--551, 2017.

\bibitem{bostrom2017strategic}
Nick Bostrom.
\newblock {Strategic implications of openness in AI development}.
\newblock {\em Global Policy}, 8(2):135--148, 2017.

\bibitem{BrookBlog2017}
Rodney Brooks.
\newblock {The Seven Deadly Sins of Predicting the Future of AI}, 2017.
\newblock
  [https://rodneybrooks.com/the-seven-deadly-sins-of-predicting-the-future-of-ai/;
  Online posted 7-September-2017].

\bibitem{brown2018superhuman}
Noam Brown and Tuomas Sandholm.
\newblock Superhuman ai for heads-up no-limit poker: Libratus beats top
  professionals.
\newblock {\em Science}, 359(6374):418--424, 2018.

\bibitem{brown2019superhuman}
Noam Brown and Tuomas Sandholm.
\newblock {Superhuman AI for multiplayer poker}.
\newblock {\em Science}, page eaay2400, 2019.

\bibitem{cave2019hopes}
Stephen Cave and Kanta Dihal.
\newblock Hopes and fears for intelligent machines in fiction and reality.
\newblock {\em Nature Machine Intelligence}, 1(2):74, 2019.

\bibitem{cave2018ai}
Stephen Cave and Se{\'a}n {{\'O}h{\'E}igeartaigh}.
\newblock {An AI Race for Strategic Advantage: Rhetoric and Risks}.
\newblock In {\em AAAI/ACM Conference on Artificial Intelligence, Ethics and
  Society}, pages 36--40, 2018.

\bibitem{Collingridge1980}
David Collingridge.
\newblock {\em The social control of technology}.
\newblock New York : St. Martin's Press, 1980.

\bibitem{Fudenberg2005}
D.~Fudenberg and L.~A. Imhof.
\newblock Imitation processes with small mutations.
\newblock {\em Journal of Economic Theory}, 131:251--262, 2005.

\bibitem{FLI_letter2015}
{Future of Life Institute}.
\newblock {Autonomous Weapons: An Open Letter from AI \& Robotics Researchers}.
\newblock Technical report, Future of Life Institute, Cambridge, MA, 2015.

\bibitem{FLI_signatories_todate}
{Future of Life Institute}.
\newblock Lethal autonomous weapons pledge.
\newblock https://futureoflife.org/lethal-autonomous-weapons-pledge/, 2019.

\bibitem{GokhalePNAS2010}
Chaitanya~S. Gokhale and Arne Traulsen.
\newblock {Evolutionary games in the multiverse}.
\newblock {\em Proc. Natl. Acad. Sci. U.S.A.}, 107(12):5500--5504, March 2010.

\bibitem{grace2018will}
Katja Grace, John Salvatier, Allan Dafoe, Baobao Zhang, and Owain Evans.
\newblock {When will AI exceed human performance? Evidence from AI experts}.
\newblock {\em Journal of Artificial Intelligence Research}, 62:729--754, 2018.

\bibitem{key:hanetalAdaptiveBeh}
T.~A. Han, L.~M. Pereira, and F.~C. Santos.
\newblock Intention recognition promotes the emergence of cooperation.
\newblock {\em Adaptive Behavior}, 19(3):264--279, 2011.

\bibitem{Han:2014tl}
The~Anh Han, Lu{\'\i}s~Moniz Pereira, and Tom Lenaerts.
\newblock {Avoiding or Restricting Defectors in Public Goods Games?}
\newblock {\em J. Royal Soc Interface}, 12(103):20141203, 2015.

\bibitem{hanAIES2019}
The~Anh Han, Lu{\'\i}s~Moniz Pereira, and Tom Lenaerts.
\newblock {Modelling and Influencing the AI Bidding War: A Research Agenda}.
\newblock In {\em Proceedings of the AAAI/ACM conference AI, Ethics and
  Society}, pages 5--11, 2019.

\bibitem{key:Hofbauer1998}
J.~Hofbauer and K.~Sigmund.
\newblock {\em Evolutionary Games and Population Dynamics}.
\newblock Cambridge University Press, 1998.

\bibitem{key:imhof2005}
L.~A. Imhof, D.~Fudenberg, and Martin~A. Nowak.
\newblock Evolutionary cycles of cooperation and defection.
\newblock {\em Proc. Natl. Acad. Sci. U.S.A.}, 102:10797--10800, 2005.

\bibitem{jobin2019global}
Anna Jobin, Marcello Ienca, and Effy Vayena.
\newblock {The global landscape of AI ethics guidelines}.
\newblock {\em Nature Machine Intelligence}, pages 1--11, 2019.

\bibitem{kandori:1993aa}
M.~Kandori, G.~J. Mailath, and R.~Rob.
\newblock Learning, mutation, and long run equilibria in games.
\newblock {\em Econometrica}, 61:29--56, 1993.

\bibitem{Karlin:book:1975}
S.~Karlin and H.~E. Taylor.
\newblock {\em A First Course in Stochastic Processes}.
\newblock Academic Press, New York, 1975.

\bibitem{MontrealDec2018}
{Montreal Declaration}.
\newblock {The Montreal Declaration for the Responsible Development of
  Artificial Intelligence Launched}.
\newblock
  https://www.canasean.com/the-montreal-declaration-for-the-responsible-development-of-artificial-intelligence-launched/,
  2018.

\bibitem{nowak:2006bo}
M.~A. Nowak.
\newblock {\em Evolutionary Dynamics: Exploring the Equations of Life}.
\newblock Harvard University Press, Cambridge, MA, 2006.

\bibitem{key:novaknature2004}
M.~A. Nowak, A.~Sasaki, C.~Taylor, and D.~Fudenberg.
\newblock Emergence of cooperation and evolutionary stability in finite
  populations.
\newblock {\em Nature}, 428:646--650, 2004.

\bibitem{pamlin2015global}
Dennis Pamlin and Stuart Armstrong.
\newblock Global challenges: 12 risks that threaten human civilization.
\newblock {\em Global Challenges Foundation, Stockholm}, 2015.

\bibitem{PwC2017}
PwC.
\newblock Sizing the prize: What's the real value of ai for your business and
  how can you capitalise?
\newblock Technical report, PwC, London, United Kingdom, 2017.

\bibitem{russell2015ethics}
Stuart Russell, S~Hauert, R~Altman, and M~Veloso.
\newblock Ethics of artificial intelligence.
\newblock {\em Nature}, 521(7553):415--416, 2015.

\bibitem{schubert2019psychology}
S~Schubert, L~Caviola, and N~Faber.
\newblock The psychology of existential risk: Moral judgments about human
  extinction.
\newblock {\em Scientific Reports}, 9(15100), 2019.

\bibitem{key:Sigmund_selfishnes}
Karl Sigmund.
\newblock {\em The Calculus of Selfishness}.
\newblock Princeton University Press, 2010.

\bibitem{silver2018general}
David Silver, Thomas Hubert, Julian Schrittwieser, Ioannis Antonoglou, Matthew
  Lai, Arthur Guez, Marc Lanctot, Laurent Sifre, Dharshan Kumaran, Thore
  Graepel, et~al.
\newblock A general reinforcement learning algorithm that masters chess, shogi,
  and go through self-play.
\newblock {\em Science}, 362(6419):1140--1144, 2018.

\bibitem{silver2017mastering}
David Silver, Julian Schrittwieser, Karen Simonyan, Ioannis Antonoglou, Aja
  Huang, Arthur Guez, Thomas Hubert, Lucas Baker, Matthew Lai, Adrian Bolton,
  et~al.
\newblock Mastering the game of go without human knowledge.
\newblock {\em Nature}, 550(7676):354, 2017.

\bibitem{sotala2014responses}
Kaj Sotala and Roman~V Yampolskiy.
\newblock {Responses to catastrophic AGI risk: a survey}.
\newblock {\em Physica Scripta}, 90(1):018001, 2014.

\bibitem{steels2018barcelona}
Luc Steels and Ramon Lopez~de Mantaras.
\newblock The barcelona declaration for the proper development and usage of
  artificial intelligence in europe.
\newblock {\em AI Communications}, (Preprint):1--10, 2018.

\bibitem{szolnoki2013correlation}
Attila Szolnoki and Matja{\v{z}} Perc.
\newblock Correlation of positive and negative reciprocity fails to confer an
  evolutionary advantage: Phase transitions to elementary strategies.
\newblock {\em Physical Review X}, 3(4):041021, 2013.

\bibitem{taddeo2018regulate}
Mariarosaria Taddeo and Luciano Floridi.
\newblock Regulate artificial intelligence to avert cyber arms race.
\newblock {\em Nature}, 556(7701):296--298, 2018.

\bibitem{traulsen2006}
A.~Traulsen, M.~A. Nowak, and J.~M. Pacheco.
\newblock Stochastic dynamics of invasion and fixation.
\newblock {\em Phys. Rev. E}, 74:11909, 2006.

\bibitem{van2012emergence}
Sven Van~Segbroeck, Jorge~M. Pacheco, Tom Lenaerts, and Francisco~C. Santos.
\newblock Emergence of fairness in repeated group interactions.
\newblock {\em Physical review letters}, 108(15):158104, 2012.

\bibitem{vinuesa2019role}
Ricardo Vinuesa, Hossein Azizpour, Iolanda Leite, Madeline Balaam, Virginia
  Dignum, Sami Domisch, Anna Fell{\"a}nder, Simone Langhans, Max Tegmark, and
  Francesco~Fuso Nerini.
\newblock The role of artificial intelligence in achieving the sustainable
  development goals.
\newblock {\em Nature Communications}, 11(233), 2020.

\end{thebibliography}

\includepdf[pages=-]{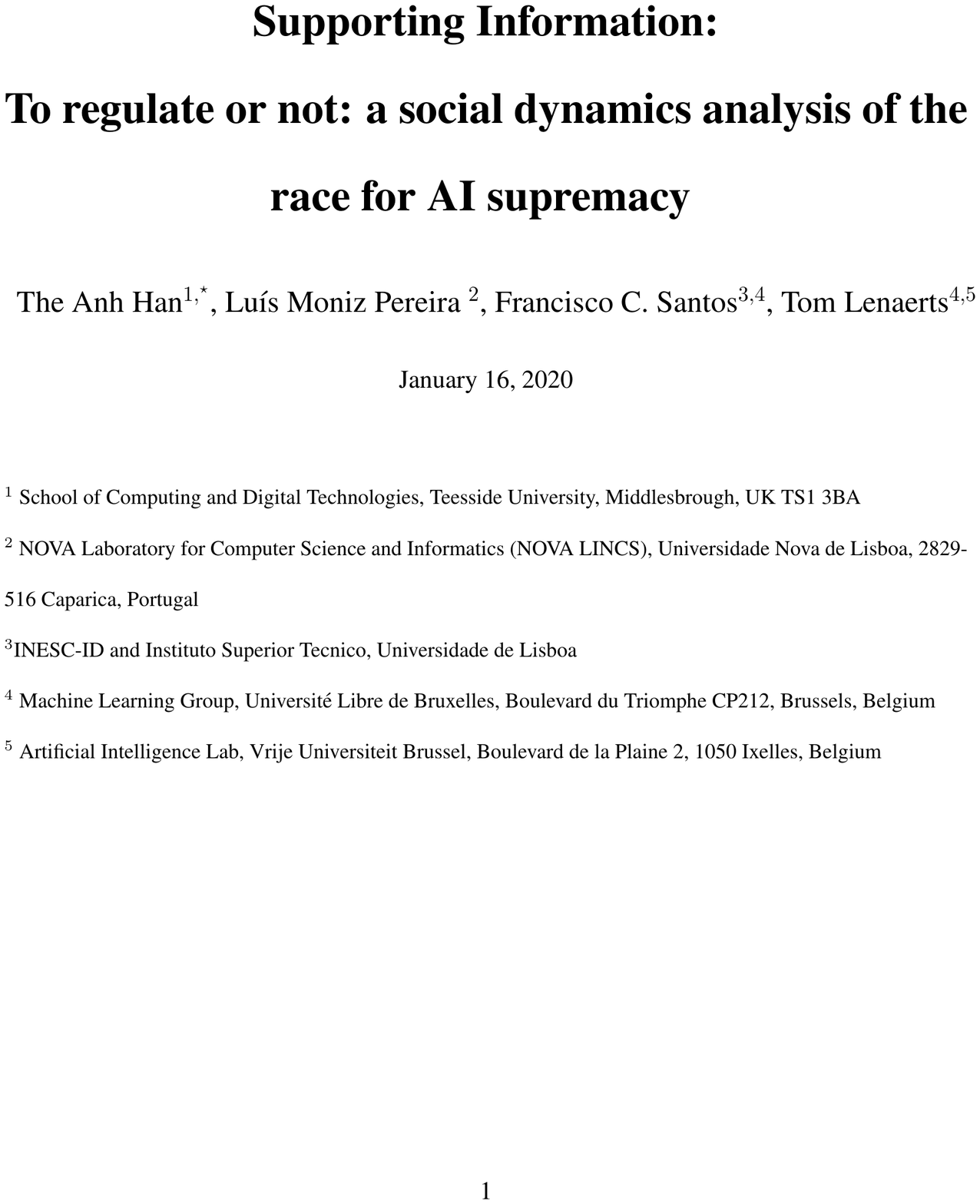}

\end{document}